\PassOptionsToPackage{unicode}{hyperref}
\PassOptionsToPackage{hyphens}{url}
\PassOptionsToPackage{dvipsnames,svgnames,x11names}{xcolor}
\documentclass[
  12pt]{article}
\usepackage{xcolor}
\usepackage{amsmath,amssymb}
\usepackage{iftex}
\ifPDFTeX
  \usepackage[T1]{fontenc}
  \usepackage[utf8]{inputenc}
  \usepackage{textcomp} 
\else 
  \usepackage{unicode-math} 
  \defaultfontfeatures{Scale=MatchLowercase}
  \defaultfontfeatures[\rmfamily]{Ligatures=TeX,Scale=1}
\fi
\usepackage{lmodern}
\ifPDFTeX\else
\fi
\IfFileExists{upquote.sty}{\usepackage{upquote}}{}
\IfFileExists{microtype.sty}{
  \usepackage[]{microtype}
  \UseMicrotypeSet[protrusion]{basicmath} 
}{}
\makeatletter
\@ifundefined{KOMAClassName}{
  \IfFileExists{parskip.sty}{%
    \usepackage{parskip}
  }{
    \setlength{\parindent}{0pt}
    \setlength{\parskip}{6pt plus 2pt minus 1pt}}
}{
  \KOMAoptions{parskip=half}}
\makeatother
\makeatletter
\ifx\paragraph\undefined\else
  \let\oldparagraph\paragraph
  \renewcommand{\paragraph}{
    \@ifstar
      \xxxParagraphStar
      \xxxParagraphNoStar
  }
  \newcommand{\xxxParagraphStar}[1]{\oldparagraph*{#1}\mbox{}}
  \newcommand{\xxxParagraphNoStar}[1]{\oldparagraph{#1}\mbox{}}
\fi
\ifx\subparagraph\undefined\else
  \let\oldsubparagraph\subparagraph
  \renewcommand{\subparagraph}{
    \@ifstar
      \xxxSubParagraphStar
      \xxxSubParagraphNoStar
  }
  \newcommand{\xxxSubParagraphStar}[1]{\oldsubparagraph*{#1}\mbox{}}
  \newcommand{\xxxSubParagraphNoStar}[1]{\oldsubparagraph{#1}\mbox{}}
\fi
\makeatother

\usepackage{longtable,booktabs,array}
\usepackage{calc} 
\usepackage{etoolbox}
\makeatletter
\patchcmd\longtable{\par}{\if@noskipsec\mbox{}\fi\par}{}{}
\makeatother
\IfFileExists{footnotehyper.sty}{\usepackage{footnotehyper}}{\usepackage{footnote}}
\makesavenoteenv{longtable}
\usepackage{graphicx}
\makeatletter
\newsavebox\pandoc@box
\newcommand*\pandocbounded[1]{
  \sbox\pandoc@box{#1}%
  \Gscale@div\@tempa{\textheight}{\dimexpr\ht\pandoc@box+\dp\pandoc@box\relax}%
  \Gscale@div\@tempb{\linewidth}{\wd\pandoc@box}%
  \ifdim\@tempb\p@<\@tempa\p@\let\@tempa\@tempb\fi
  \ifdim\@tempa\p@<\p@\scalebox{\@tempa}{\usebox\pandoc@box}%
  \else\usebox{\pandoc@box}%
  \fi%
}
\def\fps@figure{htbp}
\makeatother

\usepackage[]{natbib}
\usepackage{booktabs}
\usepackage{longtable}
\usepackage{array}
\usepackage{multirow}
\usepackage{wrapfig}
\usepackage{float}
\usepackage{colortbl}
\usepackage{pdflscape}
\usepackage{tabu}
\usepackage{threeparttable}
\usepackage{threeparttablex}
\usepackage[normalem]{ulem}
\usepackage{makecell}
\usepackage{xcolor}
\usepackage{xcolor}
\usepackage{float}
\usepackage{amsmath}
\usepackage{ulem}
\AtBeginDocument{}
\makeatletter
\@ifpackageloaded{caption}{}{\usepackage{caption}}
\AtBeginDocument{%
\ifdefined\contentsname
  \renewcommand*\contentsname{Table of contents}
\else
  \newcommand\contentsname{Table of contents}
\fi
\ifdefined\listfigurename
  \renewcommand*\listfigurename{List of Figures}
\else
  \newcommand\listfigurename{List of Figures}
\fi
\ifdefined\listtablename
  \renewcommand*\listtablename{List of Tables}
\else
  \newcommand\listtablename{List of Tables}
\fi
\ifdefined\figurename
  \renewcommand*\figurename{Figure}
\else
  \newcommand\figurename{Figure}
\fi
\ifdefined\tablename
  \renewcommand*\tablename{Table}
\else
  \newcommand\tablename{Table}
\fi
}
\@ifpackageloaded{float}{}{\usepackage{float}}
\floatstyle{ruled}
\@ifundefined{c@chapter}{\newfloat{codelisting}{h}{lop}}{\newfloat{codelisting}{h}{lop}[chapter]}
\floatname{codelisting}{Listing}

\makeatother
\makeatletter
\@ifpackageloaded{caption}{}{\usepackage{caption}}
\@ifpackageloaded{subcaption}{}{\usepackage{subcaption}}
\makeatother
\usepackage{bookmark}
\IfFileExists{xurl.sty}{\usepackage{xurl}}{} 
\makeatletter
\@ifundefined{xmpquote}{\newcommand{\xmpquote}[1]{#1}}{}
\makeatother
\hypersetup{
  pdftitle={Engaging Students With Statistics Through Choice of Real Data Context on Homework},
  pdfauthor={Catalina M. Medina; Mine Doğucu},
  pdfkeywords={\xmpquote{statistics education}, \xmpquote{data science
education}, \xmpquote{real data}, \xmpquote{datasets with
context}, \xmpquote{student autonomy}, \xmpquote{student interest}},
  colorlinks=true,
  linkcolor={blue},
  filecolor={Maroon},
  citecolor={Blue},
  urlcolor={Blue},
  pdfcreator={LaTeX via pandoc}}

\begin{document}

\def\spacingset#1{\renewcommand{\baselinestretch}%
{#1}\small\normalsize} \spacingset{1}


\title{\bf Engaging Students With Statistics Through Choice of Real Data
Context on Homework}
\author{
Catalina M. Medina\thanks{Medina has been supported by NSF IIS award
\#2123366. Medina completed an earlier part of this work in the
Department of Statistics at the University of California, Irvine.}\\
\\
\parbox[t]{0.9\textwidth}{\centering California State University Channel
Islands}\\[0.5em]
and\\[0.5em]
Mine Doğucu\thanks{Doğucu has been supported by NSF IIS award \#2123366.
Doğucu completed an earlier part of this work in the Department of
Statistics at the University of California, Irvine.}\\
\parbox[t]{0.9\textwidth}{\centering Department of Statistics}\\
\parbox[t]{0.9\textwidth}{\centering Harvard University}\\[0.5em]
}
\maketitle

\bigskip
\bigskip
\begin{abstract}
Statistics educators recommend teaching with real data with relevant
contexts, but defining relevancy is challenging and varies by student.
We investigated whether providing student choice of data context
increases engagement through a quasi-experiment in two sections of an
introductory probability and statistics course at a large public
university (n=65 consenting students). Sections alternated as treatment
and control: during their treatment, students chose weekly homework from
three similar instructor-provided options varying by data context;
during control weeks, they received randomly assigned contexts. We found
no significant difference in homework grades between treatment and
control conditions. However, thematic analysis revealed students with
choice reported enhanced engagement and motivation, greater appreciation
for statistics' real-world value, and increased autonomy. Students
overwhelmingly preferred contexts relevant to their interests,
experiences, daily lives, and career paths---though preferences varied
considerably across individuals. Based on these findings, we provide
four recommendations for statistics educators: (1) use real data with
authentic contexts, (2) select contexts students care about, (3)
incorporate variety across data contexts, and (4) consider choice as a
pedagogical tool.
\end{abstract}

\noindent%
\vfill

\newpage
\spacingset{1.9} 

\section{Introduction}\label{sec-introduction}

Among the statistics education community, there are recommendations to
use real data when teaching statistics. As Guidelines for Assessment and
Instruction in Statistics Education (GAISE) College Report
\citeyearpar{gaise2016} states in recommendation 3, instructors should
``integrate real data with a context and a purpose.'' Using real data
with context helps students perceive statistics as real and relevant
\citep{neumann2013, bibby2003, brearley2018, cobb1997, cobb1992} and
engage with the discipline \citep{hall2011}. Numerous recommendations
exist on how to teach with real data and where to find it
\citep{hall2011, singer1990}. We believe statistics should be taught as
an investigative tool to learn from data about the world.

We present Table~\ref{tbl-data-types} to clarify what we mean by real
data with a context. \emph{Made-up data} is artificially created and
easiest for instructors to generate but provides no real information.
\emph{Empirically informed data}, data simulated from real statistics or
parameter estimates, mimic real phenomena but cannot provide reliable
real-world implications. \emph{Real data} is generated from real-world
events or observations. Often we tailor real data to demonstrate
specific concepts by dropping data points, but this can remove meaning
since data no longer reflects what actually occurred. We distinguish
tailoring from cleaning, where cleaning converts data to readily usable
form. Real raw data provide exposure to data encountered in real
practice but may require substantial wrangling. Regardless of data type,
without context there is no information to be gained from statistical
analysis. This work focuses on real cleaned data with context.

\begin{table}

\caption{\label{tbl-data-types}Data context types with some pros and
cons identified.}

\centering{

\centering
\begin{tabular}[t]{>{\centering\arraybackslash}p{1.6 in}>{\centering\arraybackslash}p{0.7 in}>{\centering\arraybackslash}p{0.7 in}>{\centering\arraybackslash}p{0.7 in}>{\centering\arraybackslash}p{0.7 in}>{\centering\arraybackslash}p{0.7 in}}
\toprule
\multicolumn{4}{c}{ } & \multicolumn{2}{c}{Real-world implications} \\
\cmidrule(l{3pt}r{3pt}){5-6}
Data Type & Easily Obtained & Readily Usable & Realistic Practice & With Context & Without Context\\
\midrule
\cellcolor{gray!10}{Simulated (made up)} & \cellcolor{gray!10}{$\checkmark$} & \cellcolor{gray!10}{$\checkmark$} & \cellcolor{gray!10}{} & \cellcolor{gray!10}{} & \cellcolor{gray!10}{}\\
Simulated (empirically informed) & $\checkmark$ & $\checkmark$ &  &  & \\
\cellcolor{gray!10}{Real (tailored)} & \cellcolor{gray!10}{} & \cellcolor{gray!10}{$\checkmark$} & \cellcolor{gray!10}{} & \cellcolor{gray!10}{} & \cellcolor{gray!10}{}\\
Real (cleaned) &  & $\checkmark$ &  & $\checkmark$ & \\
\cellcolor{gray!10}{Real (raw)} & \cellcolor{gray!10}{} & \cellcolor{gray!10}{} & \cellcolor{gray!10}{$\checkmark$} & \cellcolor{gray!10}{$\checkmark$} & \cellcolor{gray!10}{}\\
\bottomrule
\end{tabular}

}

\end{table}%

When an instructor decides they want to use real data, the natural
question is what real data contexts should they use? If an instructor
decides they want to use real data with a context relevant to their
students' daily lives, there is the challenge of deciding what is
relevant to their students. \citet{sole2017} used data about coffee
sales and \citet{gould2010} recommends using data about music, social
media, maps, exercise, and data from personal devices.
\citet{gunderman2020} proposed using popular culture to engage learners
in spatial data management skills, with examples applied to topics
learners typically see as uninteresting. \citet{ludlow2002} proposed
using data from the instructor's own faculty evaluations to engage
students. \citet{brooks2021} tried to achieve relevance in their massive
open online course by using weather data from the student's region of
residence, but their findings led them to suggest using data relevant to
the students' career identities instead. In addition to recommending
data relevant to students' daily life, \citet{neumann2013} also
recommended using data relevant to a student's major.
\citet{brearley2018} discuss using research articles relevant to
students' fields. Using data relevant to students' career goals may help
bridge the gap between what teachers perceive as relevant to a student's
life and what the student perceives as relevant. These recommendations
could be helpful in a class where students are from the same or similar
majors, but maybe less helpful in a class with students from a diverse
set of majors.

Despite the efforts of teachers to find data with compelling context,
not every student will engage with every data context. Therefore there
are recommendations to use a diverse set of real data contexts
\citep{gaise2016}, with effort made to engage students with some subsets
of the data contexts investigated in the course. Alternatively, some
suggest engaging students with real data by using data collected on the
students. \citet{schneiter2023} suggest that large class sizes may be
especially poised to benefit from data collection on students by
demonstrating concepts like sampling distributions in addition to
engaging students. \citet{hulsizer2009} investigated the importance and
psychological pros and cons of using data collected on your students.

Another pathway is allowing students to choose data contexts, either in
assignments \citep{boger2001} or projects \citep{cetinkaya-rundel2022}.
\citet{boger2001} compared effects by having one section investigate
instructor-provided data while another had free choice across four
assignments. Interestingly, students given autonomy reported
significantly decreased perceived usefulness of statistics in their
field. \citet{boger2001} posited that this decrease could be due to
increased effort required for students to choose and collect their own
data.

While there have been many recommendations for data contexts to use,
there is little empirical evidence for certain contexts over others.
Even among the common recommendations of using data relevant to
students, there is little systematic study of what data students find
relevant to themselves. One thing our study sought to answer is: what do
our students perceive as relevant to themselves, and why? Investigating
this question can provide ideas and support for what could be relevant
to your students. More broadly, we hoped to investigate preferences for
and impacts of student-chosen data context to better understand what
data contexts we should be selecting as instructors, why it may or may
not be worth the effort, and if we should give students some autonomy in
selection of context.

We hypothesized that students having the autonomy to choose the data
context that they would like to work on from multiple data contexts
provided to them may lead students to be more engaged with the material.
We investigated this through a quasi-experiment. In one section of a
course we allowed students to choose their weekly homework assignment
from three similar instructor-provided options that varied by the data
context, while the other course section did not receive a choice. This
investigation of choice of data context was implemented at a large
public institution in two sections of an undergraduate introductory
probability and statistics course taught by a single instructor. As
statistics and data science instructors, we wanted to learn if we should
provide students with a choice of data context and what data contexts we
should be choosing for students? In this work, we investigated if given
the choice between similar homework assignments that differ by data
context (1) do students achieve higher grades relative to students not
given choice, (2) how do students perceive having a choice of data
context, (3) what characteristics of the data contexts led students to
choose one option over the others? We share the homework assignments we
developed and the themes we identified for research questions (2) and
(3) in an OSF repository located at {\url{https://osf.io/54rt2/}}.

\section{Methods}\label{sec-methods}

\subsection{Study Setting}\label{study-setting}

The quasi-experiment was implemented in two sections of an undergraduate
class at a large public research university in winter of 2025. The class
was Introduction to Probability and Statistics for Computer Science,
totaling 180 active students across the two sections. The authors do not
have consent to report demographics for enrolled students, but priority
enrollment into the course is given to computer science and engineering
majors. The first half of the class was mainly an introduction to
probability, but it did cover some statistical concepts like sampling
and study types, while the last half of the ten-week course was mainly
an introduction to statistics. We implemented our intervention in weeks
six through nine of the ten-week course. Intervention week 1 covered
inference for a single proportion. The next week covered inference for
two proportions and inference for a single mean. Intervention week 3
covered inference for two means and introduction to linear regression.
The final intervention week continued regression and covered maximum
likelihood estimates. The only prerequisite to the course was a
single-variable calculus course. There were 65 students who provided
consent for their data to be used for research purposes, 33 from section
A and 32 from section B.

This study involving human participants was conducted in accordance with
the ethical standards of the university's Institutional Review Board
(IRB). Under the `Exempt Self-Determination' policy, formal IRB review
and approval were not required; however, the study was registered with
the IRB (Protocol \#6666). Informed consent was obtained from all
individual participants included in the study prior to their
participation. The students were not provided any incentive to
participate (e.g., extra credit or monetary compensation), and were
informed that their professor would not know who consented to be part of
the study until after final grades were submitted.

For each week of the quasi-experiment, each student in the treatment
section was able to choose between three similar homework assignment
options, varying mainly by the data context investigated in the
homework. The students in the control section were not given options,
and were instead randomly assigned one of the three context options that
the treatment section received. Which section would first receive the
treatment, choice among assignments, was randomly assigned for the first
two weeks of the intervention period. For the last two weeks of the
intervention period the treatment and control were swapped, so each
section would ultimately experience two weeks of the intervention and
two weeks of the control condition (see
Table~\ref{tbl-experimental-design}).

\begin{table}[!htb]

\caption{\label{tbl-experimental-design}Data context quasi-experimental
design.}

\centering{

\centering
\begin{tabular}[t]{>{\raggedright\arraybackslash}p{1cm}>{\raggedright\arraybackslash}p{7cm}>{\raggedright\arraybackslash}p{7cm}}
\toprule
\multicolumn{1}{c}{ } & \multicolumn{2}{c}{Data context choices} \\
\cmidrule(l{3pt}r{3pt}){2-3}
Week & Section A & Section B\\
\midrule
\cellcolor{gray!10}{1} & \cellcolor{gray!10}{Cheating with AI, Gluten-free pasta, Home-team advantage} & \cellcolor{gray!10}{Cheating with AI}\\
2 & Cheating with AI, Electric vehicle adoption, Sleep deprivation & Sleep deprivation\\
\cellcolor{gray!10}{3} & \cellcolor{gray!10}{Typing speeds} & \cellcolor{gray!10}{Dog lifespans, Teen pregnancies, Typing speeds}\\
4 & 2003 movies & 2003 movies, Bike shares, Gorillas\\
\bottomrule
\end{tabular}

}

\end{table}%

\subsection{Characteristics of our Data
Contexts}\label{characteristics-of-our-data-contexts}

Each homework assignment was centered around one real data set, with
context provided to motivate the research questions investigated. Many
of the contexts were chosen because we believed they could be relatable
to our students. For example, ``home-team advantage'' and ``2003
movies'' could relate to students' hobbies, ``gluten-free pasta'' and
``sleep deprivation'' could relate to students' biology and daily lives,
and ``cheating with AI'' and ``typing speeds'' could relate to their
studies and life as a student, especially since the class was mainly
meant for computer science majors. The ``bike shares'' and ``electric
vehicle adoption'' datasets could both relate to their lives because
transportation with bikes and electric vehicles at the university is
fairly common.

A few of the relatable data contexts were chosen with additional
intentions. We thought that the ``sleep deprivation'' study may be of
additional interest to students because it compared people in their
state to people in a different US state, so maybe location could factor
into students' preferences. We anticipated that the ``gluten-free
pasta'' context might particularly resonate with students with gluten
intolerance, or those close to someone with gluten intolerance, because
it addresses an experience that is personally meaningful to some
students but rarely represented in course examples. The ``cheating with
AI'' was included twice because we were curious if students would prefer
a context they had previously investigated. We were curious if students
who are not interested in sports would have any negative feelings
towards the ``home-team advantage'' dataset.

The ``gorillas'' dataset was selected because we thought it was a
scientific study about a topic most students would not relate to, but
may be curious to learn about. To contrast this, we included the ``2003
movies'' in the same week because it relates to popular culture, which
might be perceived as easier to understand. We explicitly added the year
to the homework title because we were curious if students would have
opinions about the dataset containing only old movies. The ``teen
pregnancies'' dataset was chosen because we thought it was a more
serious topic than pets and typing speeds. Ironically, neither author
thought of the ``dog lifespans'' dataset as having negative associations
prior to reading the study results.

\subsection{Homework Centered Around Real Data With
Context}\label{homework-centered-around-real-data-with-context}

Each homework assignment was centered around the investigation of a real
data set with context. The homework assignments were computer-based, and
the student would either see one homework or three homework options for
that week, with the statistical topic and data context included in the
title when choice was provided.

The data context was included in the title so students would immediately
be aware of the data context, even without opening the assignment. The
assignments where students were given a choice began with ``Note: There
are three options for homework {[}number{]}, that differ by data topic
and you only need to complete one of the three assignments.'' Within the
assignment there would be a brief introduction to the data context with
either the data file, or a link to the study it came from. For example,
the sleep deprivation homework option of intervention week 2 began:
``Sleep deficiency can affect your physical, cognitive, and social
functions. Sleep deprivation has been linked to many chronic health
problems including obesity and depression. The CDC conducted a study in
2008 {[}link embedded{]} to investigate sleep deprivation among United
States residents\ldots{}''

The homework would involve an investigation of the real data within
context, with questions consistent across homework options for that week
that only varied by data context. For example, for intervention week 2
homework options, question 2 began either:

\begin{quote}
``We want to calculate a 90\% CI for the difference in proportion of
residents that experienced sleep deprivation for the past 30 days
between California and Washington\ldots{}''
\end{quote}

\begin{quote}
``We want to calculate a 90\% CI for the difference in proportion of
public high school students that used a digital device as an
unauthorized aid between 2019 and 2023\ldots{}''
\end{quote}

The real data investigation would sometimes be supplemented by
hypothetical questions within the context, but with artificial numbers
to have students explore theoretical properties, such as what happens to
the standard deviation when sample size increases. We would clearly
convey when the data was artificial so students did not have to assume.
Parallel questions from three assignments for a single week are supplied
in the Section S2 of the Supplemental Materials and the full homework
assignments are available in the OSF repository
{\url{https://osf.io/54rt2/}} in the \texttt{study-hw-assignments}
folder.

\subsection{Data Collection and
Analysis}\label{data-collection-and-analysis}

\textbf{Research question 1: do students achieve higher grades relative
to students not given choice?} First, we wanted to detect if there were
differences in weekly homework grades associated with being given a
choice of data context. We had access to all of the students' homework
grades, including those before the intervention period. Specifically, we
chose to look for differences in a student's homework grades relative to
their average pre-intervention period homework grade, associated with
having a choice of data context. We chose to investigate this with a
generalized estimating equation where the response for student \(i\) at
time \(t\) was the difference between their homework grade for that week
and their average homework grade from the first five weeks of class.
This was regressed on whether they had a choice of data context for the
homework, the student's section, and separate indicators for each
intervention week. An exchangeable correlation structure was used for
each student to control for intra-student correlation across time
points. \begin{equation}\protect\phantomsection\label{eq-hw}{
\begin{split}
&\text{Intervention Homework Grade}_{it} - \text{Average Pre-intervention Homework Grade}_{i}\\
&= \beta_0 + \beta_1 {I_{(\text{Choice})}}_{it} + \beta_2 {I_{(\text{Section B})}}_{i} + \beta_3 {I_{(\text{Week 2})}}_{t} + \beta_4 {I_{(\text{Week 3})}}_t + \beta_5 {I_{(\text{Week 4})}}_t + \epsilon_{it},\\ 
&\text{ for } i = 1, ..., 65, t = 1,..., 4.  
\end{split}
}\end{equation}

\textbf{Research question 2: how do students perceive having a choice of
data context?} We surveyed students after the intervention period to
answer research question 2. We asked ``What did you think about having
three homework assignment options to choose from that differed by data
context for two homework assignments in this class?'', since each
section only had choice of data context for two weekly homework
assignments. We used thematic analysis in an iterative process to
identify recurring subthemes among the students' responses to this
question. Both researchers independently identified subthemes, and
judged each response to be overall negative, impartial, or positive
\citep{braun2006}. When the researchers disagreed about subthemes, it
typically resulted in a new subtheme being identified, which was more
specific to that case. We then grouped the subthemes into larger themes.
This would also result in revisiting already classified responses with
the new set of subthemes available.

\textbf{Research question 3: what characteristics of the data contexts
led students to choose one assignment over the other options?} The
second goal of this work was to investigate why students chose a data
context over the other options, and what they did or did not like about
a given data context. At the end of each homework assignment the
students had a question related to the data context, which varied
depending on whether the student had a choice of data context for that
assignment. Consider intervention homework 1 and the cheating with AI
topic, which was the randomly assigned topic for section B.

The students not given a choice of data context (section B) were asked:
``Please explain in detail what you liked and/or what you did not like
about this data topic.'' Students given a choice of data context
(section A) were asked: ``Please explain in detail what led you to
choose this topic, cheating with AI, to investigate'' and ``What was it
about the other two options (gluten-free pasta and home-team advantage)
that led you not to choose them? Please explain in detail.'' These
questions also specified that there were no correct answers and these
questions were only being graded for completeness.

Thematic analysis was also used to identify subthemes and then overall
themes, similarly to the process described for research question 2. The
themes were recorded as either being specific to a context, or
nonspecific to the context (e.g., the student just chose the first
homework option on the screen).

We forgot to include the ``what you liked and/or what you did not like
about this data topic'' question in the intervention week 3 homework
assignment for the no-choice section (section A), so we do not have data
from that section for that week. If we did have these data, it would not
change the subthemes and themes we identified, but it is possible that
new subthemes could have been generated. Our final codebooks for
research questions 2 and 3 are provided in Tables S2 and S3 of the
Supplemental Materials.

\section{Results}\label{results}

Analysis of surveys and homework revealed four overarching categories
related to having a choice of real data context: (1) students' sentiment
about having a choice, (2) student-perceived qualitative outcomes, (3)
quantitative outcomes, and (4) students' desired characteristics of real
data contexts.

\subsection{Students' Sentiment About Having
Choice}\label{students-sentiment-about-having-choice}

From the post-survey we judged 2 students to have an overall negative
opinion, 18 overall impartial opinions, and 43 students to have an
overall positive opinion of having a choice of data context.

The two students with an overall negative opinion thought having a
choice ``added unnecessary difficulty'' or ``made starting the
assignment a little harder since I had to decide which topic to pick.''
This indicated that the students somewhat cared about which homework to
pick, but unfortunately they thought the ``options were still a bit
limited.'' While only two students had overall negative opinions, there
were two more comments about negative aspects of having choice. A couple
of students expressed that having their peers work on different
assignments made ``it more difficult to discuss the best way to solve
certain problems'' or ``understand the concepts if they work on
differing work.'' Another student expressed the opposite opinion on the
situation, stating ``I do like that having variety means that it's not
as easy to copy answers from other students.''

Most of the students who had overall impartial opinions expressed that
having a choice of data context had no effect on them. Some even
specifically said it ``didn't make my homework more or less enjoyable''
or ``they provide no significant benefit to my learning of the course
topics.'' A few students expressed disinterest in the offered topics,
with one saying ``the topics would only matter to me if they were
actually interesting to me, and if they're just three random topics that
all equally are not that interesting to me then there's no point for
me.'' One student even said ``the context of the analysis wasn't as
useful as the actual data was,'' which is the opposite of the
instructor's goal of working with real data with a real context on the
homework. A student who was overall impartial did acknowledge that while
they were ``ambivalent on the matter\ldots{} I do think that it could
benefit some students.''

Among all sentiments, there were students who simply appreciated having
autonomy, regardless of what that autonomy pertained to. ``I really
liked having the ability to choose. That small privilege somewhat made
the homework more pleasing.'' One student even expressed ``I found it
more interesting when I got to choose perhaps that's the illusion of
choice.'' A couple of students pointed out that having multiple options
for homework assignments provided opportunities for further studying and
practice with the assignments they did not choose.

Many of the students who expressed an overall positive sentiment
described experiencing some qualitative outcomes resulting from having
choice of data context, which will be discussed in detail in the
following section.

\subsection{Qualitative Outcomes}\label{qualitative-outcomes}

While we did not intend to investigate any associations with our
intervention other than grade differences, thematic analysis revealed
many themes of students identifying positive experiences. The following
is an exemplary response:

\begin{quote}
\emph{I thought that having more options on what context to do problems
with was pretty interesting overall, and helped motivate me to do the
homework knowing that there was a tangible and significant reason for
doing it. By that I mean that doing the homework to solve a problem I'm
interested in seems more like something I would want to do on my
off-time than something done for a grade, so I was more engaged in the
assignment.}
\end{quote}

\begin{figure}

\centering{

\includegraphics[width=6in,height=3in,alt={Three encapsulating ovals. The outer oval is titled \textquotesingle Real data from a real context\textquotesingle{} and lists \textquotesingle value of statistics\textquotesingle, \textquotesingle recognize real-world application of statistics\textquotesingle{} and learn about context from data\textquotesingle. The next inner oval is titled \textquotesingle Read data from context students are interested in\textquotesingle{} and lists \textquotesingle Desire to learn more about data context\textquotesingle{} \& \textquotesingle better experience with homework\textquotesingle. The inner most oval is titled \textquotesingle Real data from context students are interested in and chose themselves\textquotesingle{} and lists \textquotesingle Sense of autonomy\textquotesingle, and \textquotesingle looking forward to choice on future assignments\textquotesingle.}]{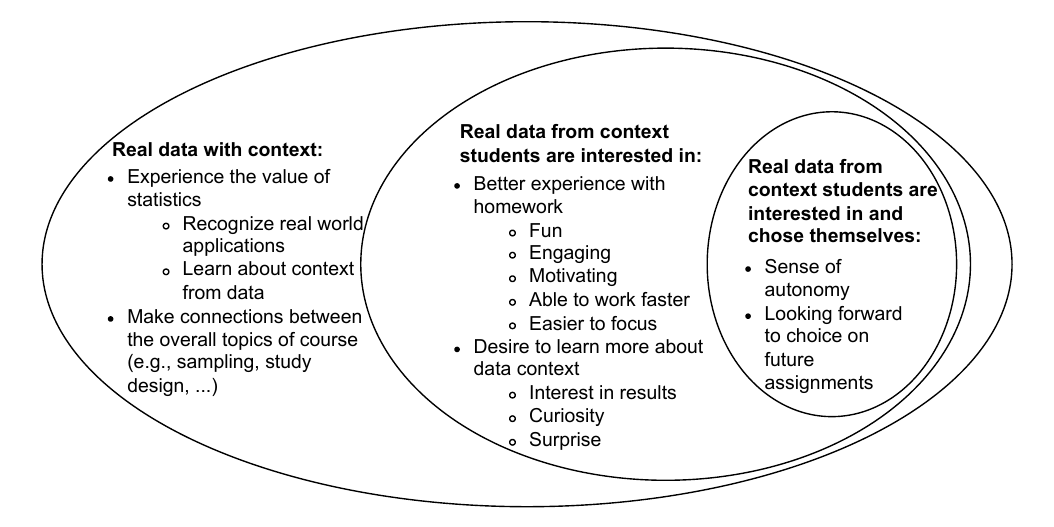}

}

\caption{\label{fig-perceived-positive-outcomes}Students' perceived
positive outcomes.}

\end{figure}

We classified students' responses as being associated with working with
real data from a real context, working with real data from a context
students are interested in, and working with real data from a context
students are interested in and chose themselves, visualized in
Figure~\ref{fig-perceived-positive-outcomes}. It is important to note
that we had a low consent rate, so the identified themes were not
necessarily representative of all of the students in the class, but we
can still learn from the themes that consistently emerged.

As a student expressed, when working through an assignment using real
data with context ``statistics homework will let the student get more
insight on the topic.'' Moreover, it ``showed that statistics is not
just some hypothetical numeric calculation, but rather something that
could actually have real-world impacts.'' Instead of instructors arguing
why statistics is useful, students were reporting that they experienced
it being useful. Furthermore, some students were even thinking beyond
the homework questions by discussing considerations of a study's design
or concerns about bias due to the sampling scheme. ``It was cool to see
survey results showing whether students admit to using AI. I just wish
the sample sizes were bigger so we'd be more confident in the
findings.'' Having a real context provides depth for students to apply
all of their statistical knowledge, not just the content they learned
that week.

More benefits were identified when the data was not only real with
context, but specifically from a context students were interested in:
``Rather than looking at the probabilities of flipping heads or rolling
a die, I enjoyed learning through this example because it has a greater
real-life impact.'' Students often expressed this led to the homework
being more fun, engaging, and/or motivating, with some saying they were
able to work faster and/or had an easier time focusing. A desire to
learn more about the data context was expressed both in why they chose a
context and in their discussion of the results they found in the
homework. ``I have a large dog at home who I love A LOT. I can never
think about his lifespan and only wish for him to live forever. However,
I chose this study in hopes of understanding more about the topic.'' At
the end of homework assignments, when we asked why they did or didn't
choose certain contexts, several students used this as an opportunity to
further describe or explore the results they found. They reasoned
whether or not the results made sense to them given their knowledge or
experience, articulated surprise at unexpected patterns, posed possible
explanations for findings, and one looked into unused variables in the
data to explore more associations.

The innermost ring of Figure~\ref{fig-perceived-positive-outcomes}
highlights the positive outcomes associated with students being able to
choose their context. Providing a choice of context was not only a
mechanism to increase the chances that students would work with a
context they are more interested in. A student said the autonomy itself
``\ldots{} allowed us to express some creativity and independence of
choice'' and another said ``that small privilege somewhat made the
homework more pleasing.'' One person said knowing they would have a
choice made them look forward to their next assignment. In further
support of students appreciating choice, a student expressed that they
enjoyed having a choice, but that ``it was annoying when it switched
from three options to one option because the other class was able to
pick more interesting topics.''

\subsection{Quantitative Outcomes}\label{quantitative-outcomes}

Using the model described in Equation~\ref{eq-hw}, we did not find
evidence of a difference between a student's homework score relative to
their average pre-intervention homework score, for a given section and
intervention week (p-value = 0.2377). Specifically, we estimate this
difference to be between -3.3 and 0.8 with 95\% confidence. This means
we are uncertain about the direction and magnitude of the change in
homework grade associated with students' ability to choose the data
context for their homework assignment, after accounting for section and
intervention week. Further model results are available in Table S1 of
the Supplemental Materials.

While a lack of an effect of choice could possibly explain these
results, we note they could also be due to other factors such as low
statistical power or a misspecified model. A alternative approach better
suited to smaller sample sizes would have been to use Bayesian modeling.
For this model, we chose an exchangeable within-student covariance
structure to account for within-student variation and the limitation of
our small sample size. A better option may be to account for an expected
difference in correlation between the two treatment homework assignments
and between the two control homework assignments, if there was more
statistical power. Finally, a challenge of modeling homework grades is
the ceiling effect. Any theoretical improvement from an already
near-perfect-scoring student would be nearly unobservable, depending on
how close the score is to the maximum.

\subsection{Students' Preferred Real Data Context
Characteristics}\label{students-preferred-real-data-context-characteristics}

\begin{figure}[!htb]

\centering{

\pandocbounded{\includegraphics[keepaspectratio]{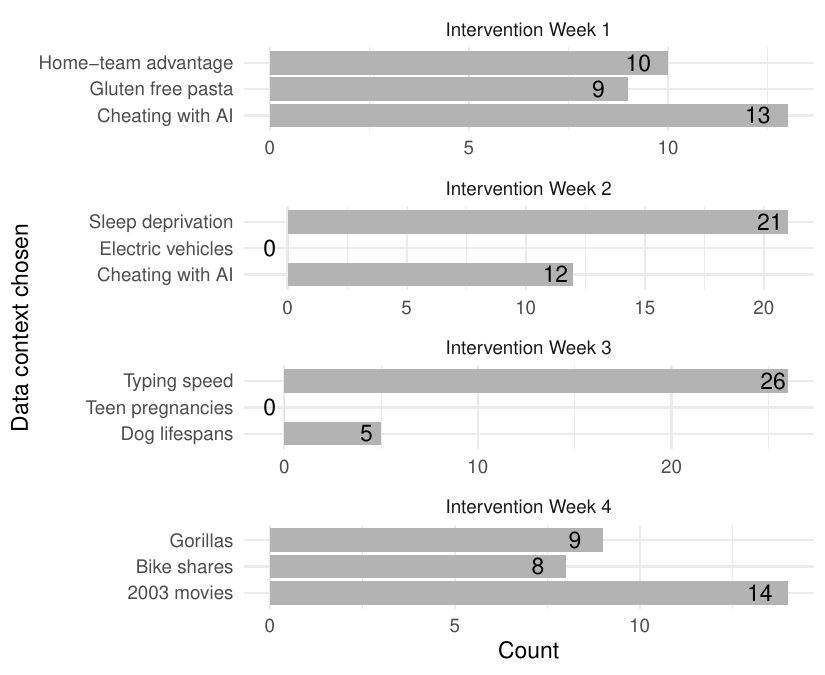}}

}

\caption{\label{fig-chosen-contexts}Number of students who chose each
data context by week, among students given choice of data context.}

\end{figure}%

Figure~\ref{fig-chosen-contexts} displays the counts of how many
students chose a given data context, among lecture A students for the
first two intervention weeks and lecture B students for the last two. No
student chose electric vehicles in intervention week 2 or teen
pregnancies in week 3. A student who chose the dog lifespan topic
instead of the teen pregnancy topic expressed ``I didn't choose teen
pregnancy since those statistics aren't very fun to learn about and are
more depressing compared to dogs.'' Additionally, very few students
chose dog lifespans in week 3.

\begin{figure}

\centering{

\includegraphics[width=6in,height=3in]{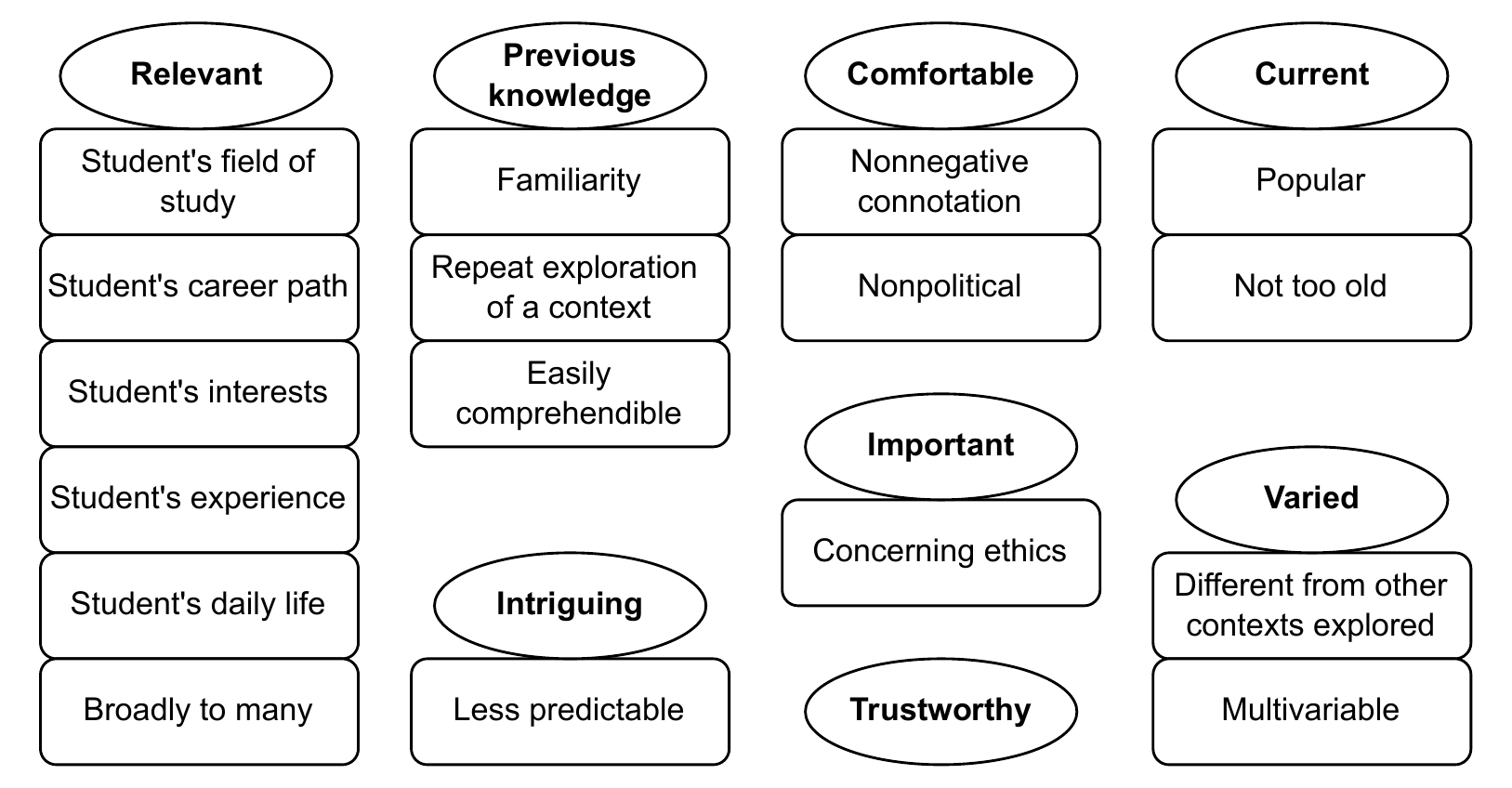}

}

\caption{\label{fig-desirable-data-context-characteristics}Desirable
data context characteristics expressed by students.}

\end{figure}

Themes of why students chose certain contexts and not others are
summarized in Figure~\ref{fig-desirable-data-context-characteristics}.
Two major takeaways are that students have conflicting opinions and they
overwhelmingly prefer contexts that are relevant to them and about which
they have some previous knowledge.

To the first point, some students liked a topic because they felt it was
important or commonly discussed, while others disliked it for those very
reasons. Some students chose the AI topic the second week because they
had chosen it for the first topic, but other students either didn't like
the context in the first place or preferred learning about a variety of
contexts. Some students found the 2003 movies data to be too old because
it was from around or before they were born while others preferred the
dataset because they liked movies more than they cared about them being
old.

People caring more about things that are relevant to them is not a new
idea, but relevance can be very difficult to identify for an instructor.
These students' responses can help by providing more specific ways
contexts can be relevant to them, such as their field of study, career
path, interests, experiences, and daily lives.

Some preferences were less obvious, such as liking a context because
they did not know what the results were likely to be beforehand or not
liking a context because they felt a lack of trust in the quality of the
data inherently due to the topic. This one arose primarily from the
cheating with AI topic where students assumed there would be
underreporting. Another less commonly discussed preferred characteristic
was for richer data that allow them to look further into associations
beyond what was explored in the homework.

\section{Discussion}\label{sec-discussion}

\subsection{Recommendations for Statistics and Data Science
Educators}\label{recommendations-for-statistics-and-data-science-educators}

\textbf{Recommendation 1: Use real data with a real context}\\
\underline{Why?} A major theme we identified was students wanting to
learn about the data context through the assignment, or expressing
interest in the results of the analysis. This is not possible without
real data from a real context, and artificial data from a real context
may just perpetuate the perception of statistics being fake or made up.
Our findings in support of the use of real data with a real context are
in line with the empirical findings of \citet{neumann2013} and the
recommendations of many in the literature, including \citet{gaise2016}.

\underline{How?} We believe statistics and data science should be taught
with real data generally, but we are not arguing against the use of
simulated data when appropriate. Simulated data can be very helpful for
exploring theoretical properties and ideas. We are arguing against the
use of artificial data in place of real data, especially when the
argument is just that using artificial data is easier for the
instructor. To facilitate this, we provide a list of resources we have
found useful for finding datasets for educational purposes, available in
Section S3 of the Supplementary Materials.

\textbf{Recommendation 2: Use data contexts students care about}\\
\underline{Why?} \citet{neumann2013} found a common theme of students
finding the use of real data engaging and motivating, and we
additionally found empirical evidence of students being further engaged
and motivated by real data with data contexts they were interested in.
Improving students' experiences when learning statistics is a great
motivator for instructors to strive to include interesting datasets for
students.

Contrary to \citet{young-saver2024}, while we did not find a change in
quantitative outcomes, we did have a lot of students saying they had a
better experience with their homework when they were more interested or
familiar with a topic. Even though there were also many students who did
not feel choice of data context made any difference in their experience,
we still think that being able to better engage some students is worth
the effort.

\underline{What?} We found several themes among students' data context
preferences that can help us learn what students care about. Students
preferred data that was relevant to them, their experiences, daily lives
or lives as students, and their academic, career, and personal
interests.

Many students avoided the dog lifespan and teen pregnancy topics,
perceiving them as negative or challenging. The dog lifespan context
inherently involves death, so we don't feel this was a misunderstanding.
However, we believe we inadequately introduced the teen pregnancy topic,
resulting in an unintended perception. When working with sensitive or
uncomfortable topics, it's important to be extra conscious of how we
introduce the data context to students.

\underline{How?} Students often associated familiarity with a topic with
easier homework. Instructors should introduce data contexts before
expecting students to engage with data. Sensitive or intimidating
contexts could be introduced prior to appearing in the assignment,
allowing for discussion around the topic and possibly alleviating
discomfort. When introducing statistical concepts students struggle
with, use familiar and easily understood data contexts to reduce the
burden of simultaneously understanding both the concept and the context.

We believe that statistics can fuel a love of science, but scientific
studies can be intimidating for both instructors and students to
understand and they run the risk of being perceived as ``just numbers.''
\citet{evans2023} recommended using a study on human-dog interaction as
a more accessible option for increasing engagement in undergraduate
statistics courses. Other scientific studies can be made accessible by
introducing them in ways connected to students. For example, we may want
to use the study ``Microbiome of Clothing Items Worn for a Single Day in
a Non-Healthcare Setting'' because it is relatively accessible with only
basic statistics used \citep{whitehead2023} and can be made relevant to
students by asking ``Think about wearing your clothes all day and then
sitting in your bed with them. What bacteria or fungi have
accumulated?'' \citet{singer1990} proposed using data from local
dissertations. We think this could be a great way to involve real data
relevant to students and it could help demystify the intimidating topics
of scientific research, graduate school, and dissertations while helping
students see what their own institution is contributing to society.

Another way to provide students with data contexts they care about is to
ask them through a survey \citep{weiland2024}, or even collect data on
the students themselves to use in class \citep{schneiter2023}. From this
feedback, a few datasets used in the class could be selected for
substitution.

We do not recommend relying solely on our own perceptions of what
students will care about. As anecdotal evidence, both investigators
predicted which data contexts would be most chosen and were wrong every
week. Despite working with our student population for a while, we were
severely incorrect about their preferences. Instead of choosing data
based on our own opinions, we can be more successful by listening to
what students say they care about.

\textbf{Recommendation 3: Teach with a variety of data contexts}\\
\underline{Why?} No single data context appeals to all students, so
limiting contexts reduces engagement opportunities. Preferences change
over time and between classes---even within this class meant for
computer science majors, some found typing speed or the cheating with AI
dataset extremely boring. While we cannot make everyone happy, variety
helps us reach more students throughout the course. Students working
with different contexts were able to see how statistics is meaningful
and applicable to real-world problems. We want students to see that the
same statistical methods are useful across many different contexts, not
just one.

\underline{How?} Reflecting on data contexts used throughout your course
can help assess variety. Getting data from diverse sources helps prevent
contexts from being too similar.

In addition to incorporating a variety of contexts, instructors can use
rich datasets with multiple variables to encourage student engagement
and exploration. The 2003 movies dataset was chosen most often in the
final intervention week, and several students described examining movies
or investigating relationships beyond those required by the assignment.
These students independently pursued questions prompted by their
interest in the data. The GAISE 2016 College Report recommends
developing multivariable thinking as a learning goal
(\citet{gaise2016}). Rich datasets complement this goal: instructors can
use selected variables for guided explorations in class or on homework
while encouraging students to investigate additional variables and
relationships based on their own curiosity.

\textbf{Recommendation 4: Consider choice as a pedagogical tool}\\
\underline{Why?} Even though we provided a variety of datasets, we still
had students who were sometimes not intrigued by any of the options. By
providing students with choices of data context, we can accommodate
different interests at the same time, instead of only engaging each
student a few times throughout the course. Additionally, students
appreciate having autonomy, even for small tasks like choosing between
similar homework assignments. It also provides students with more
resources to practice the material beyond the required assignments. As
discussed in the introduction, some instructors provide choice of data
context through a final project. Alternatively, by providing some
choices throughout the course we can hopefully engage more students
before the end of the course. A consideration to be aware of is when you
want all students to interact with a certain data context, versus when
you are okay with them having a choice.

\underline{Why not?} We recognize that it may not be feasible for all
instructors or for all assignments. Beyond the initial burden of finding
and incorporating multiple real datasets with contexts we think students
will care about, there is also the struggle of grading different
assignments. The burden of finding data can be alleviated across
multiple iterations of a course, but the burden of grading assignments
should be taken into consideration. This burden is similar to creating
and grading multiple versions of an exam. For example, the different
datasets may result in different conclusions, like a p-value of 0.0001
versus 0.045. Another example is that some data contexts may be less
familiar to students, so interpretation could be more difficult for one
context relative to another. In general, rubric-based grading can help
standardize your grading across the homework options, and these
considerations can help guide an instructor's selection of data context
options for a particular assignment to try and achieve moderate
consistency across assignment options.

\subsection{Limitations and Future
Directions}\label{limitations-and-future-directions}

Our sample was limited by being a major-specific course taught by a
single professor and the random assignment of treatment was done by
section not at an individual level, so it was only a quasi-experiment.
The low student consent rate and the resulting nonrepresentative sample
prevent us from generalizing our results broadly.

There is the potential for a novelty effect in our results, where a
student may temporarily experience positive outcomes because the
intervention is something new, not necessarily because it will continue
to be effective. Each section only experienced two intervention weeks,
so our study period was too short to provide insight into this. If
present, a novelty effect could support the occasional, rather than
consistent, use of choice as a pedagogical tool. A future study would be
useful to evaluate this possibility by implementing choice over a longer
period of time, like the entire course.

A carry-over effect is also possible in this study, and we saw some
evidence to suggest how it could occur from the student who expressed
annoyance when choice was taken away and given to the other section. In
this case, the carry-over effect could be negative after choice is taken
away, but we cannot determine whether this reaction affected the
student's engagement or performance. There is potential for a positive
carry-over effect, where a student remains more engaged with the course
due to previously experiencing the intervention, but we currently cannot
conclude if this was present. If present, it could contribute to the
null result we observed in the quantitative study.

An additional consideration is self-selection bias, which would limit
generalizability of the quantitative findings. Self-selection bias could
have resulted in responses from students who are more engaged with the
course than those who did not consent to participate, limiting the range
of perspectives represented in these data. Although qualitative findings
are evaluated in terms of transferability rather than generalizability,
this potential omission should be considered when interpreting the
identified themes. Nevertheless, the differing experiences and
preferences expressed by participating students suggest that teaching
with a variety of contexts may increase your chances of engaging more
students.

\subsection{Conclusions}\label{conclusions}

This was only an initial investigative study and will require further
investigation, but our identified themes provide empirically supported
avenues for future targeted investigations. This work also provides key
considerations for instructors when selecting the data they teach with
and a starting point for identifying data contexts that may interest
students and, in turn, support greater engagement.

\section{Data Availability Statement}\label{data-availability-statement}

The data-cleaning scripts, the themes identified in the qualitative
analysis, and the educational materials (homework assignments) developed
as interventions for this study are openly available on the Open Science
Framework repository located at {\url{https://osf.io/54rt2/}}.
De-identified quantitative data may be made available by the
corresponding author upon reasonable request for research purposes,
subject to an appropriate data-use agreement.

\section{IRB Statement}\label{irb-statement}

This study involving human participants was conducted in accordance with
the ethical standards of the university's Institutional Review Board
(IRB). Under the `Exempt Self-Determination' policy, formal IRB review
and approval were not required; however, the study was registered with
the IRB (Protocol \#6666). Informed consent was obtained from all
individual participants included in the study prior to their
participation. The students were not provided any incentive to
participate (e.g., extra credit or monetary compensation), and were
informed that their professor would not know who consented to be part of
the study until after final grades were submitted. Raw qualitative
responses are not publicly available because they contain descriptive
details that could potentially compromise participant anonymity.

\section{AI statement}\label{ai-statement}

No AI was used in the planning or execution of this study or in the
original drafting of this manuscript. GPT-5.6 Sol Light and Claude Sonet
5 was used in the revision process to check for typos and grammar, to
help with LaTex formatting, and to identify whether all of the requested
changes had been addressed and/or implemented.

\section{Acknowledgements}\label{acknowledgements}

Both authors have changed institutions since the start of this project.
{Dr.~Doğucu} did the majority of this work while working at {the
University of California, Irvine}. {Dr.~Medina} did an earlier part of
this work at {the University of California, Irvine}.

\section{Disclosure Statement}\label{disclosure-statement}

No potential competing interest was reported by the authors.

\section*{References}\label{references}
\addcontentsline{toc}{section}{References}

\renewcommand{\bibsection}{}
\bibliography{references.bib}

@book{hulsizer2009,
	title = {A Guide to Teaching Statistics: Innovations and Best Practices},
	author = {Hulsizer, Michael R. and Woolf, Linda M.},
	year = {2009},
	month = {01},
	date = {2009-01-30},
	publisher = {John Wiley \& Sons},
	note = {Google-Books-ID: Z5oNcfVeEN4C},
	langid = {en}
}

@article{gould2010,
	title = {Statistics and the Modern Student},
	author = {Gould, Robert},
	year = {2010},
	date = {2010},
	journal = {International Statistical Review},
	pages = {297--315},
	volume = {78},
	number = {2},
	doi = {10.1111/j.1751-5823.2010.00117.x},
	url = {\url{https://doi.org/10.1111/j.1751-5823.2010.00117.x}},
	langid = {es}
}

@article{cobb1997,
	title = {Mathematics, Statistics, and Teaching},
	author = {Cobb, George W. and Moore, David S.},
	year = {1997},
	month = {11},
	date = {1997-11},
	journal = {The American Mathematical Monthly},
	pages = {801--823},
	volume = {104},
	number = {9},
	doi = {10.1080/00029890.1997.11990723},
	url = {\url{https://doi.org/10.1080/00029890.1997.11990723}},
	langid = {en}
}

@article{neumann2013,
	title = {Using Real-Life Data When Teaching Statistics: Student Perceptions of This Strategy in an Introductory Statistics Course},
	author = {Neumann, David L. and Hood, Michelle and Neumann, Michelle M.},
	year = {2013},
	month = {11},
	date = {2013-11-29},
	journal = {Statistics Education Research Journal},
	pages = {59--70},
	volume = {12},
	number = {2},
	doi = {10.52041/serj.v12i2.304},
	url = {\url{https://doi.org/10.52041/serj.v12i2.304}},
	note = {Number: 2},
	langid = {en}
}

@article{brooks2021,
	title = {Towards Culturally Relevant Personalization at Scale: Experiments with Data Science Learners},
	author = {Brooks, Christopher and Quintana, Rebecca M. and Choi, Heeryung and Quintana, Chris and NeCamp, Timothy and Gardner, Joshua},
	year = {2021},
	month = {09},
	date = {2021-09-01},
	journal = {International Journal of Artificial Intelligence in Education},
	pages = {516--537},
	volume = {31},
	number = {3},
	doi = {10.1007/s40593-021-00262-2},
	url = {\url{https://doi.org/10.1007/s40593-021-00262-2}},
	langid = {en}
}

@article{singer1990,
	title = {Improving the Teaching of Applied Statistics: Putting the Data Back into Data Analysis},
	author = {Singer, Judith D. and Willett, John B.},
	year = {1990},
	month = {08},
	date = {1990-08-01},
	journal = {The American Statistician},
	pages = {223--230},
	volume = {44},
	number = {3},
	doi = {10.1080/00031305.1990.10475726},
	url = {\url{https://doi.org/10.1080/00031305.1990.10475726}},
	publisher = {ASA Website}
}

@article{schneiter2023,
	title = {Leveraging the {\textquotedblleft}Large{\textquotedblright} in Large Lecture Statistics Classes},
	author = {Schneiter, Kady and Hadfield, KimberLeigh Felix and Clements, Jenny Lee},
	year = {2023},
	month = {05},
	date = {2023-05-04},
	journal = {Journal of Statistics and Data Science Education},
	pages = {173--178},
	volume = {31},
	number = {2},
	doi = {10.1080/26939169.2022.2099488},
	url = {\url{https://doi.org/10.1080/26939169.2022.2099488}},
	publisher = {Taylor \& Francis}
}

@article{sole2017,
	title = {What's Brewing? A Statistics Education Discovery Project},
	author = {Sole, Marla A. and Weinberg, Sharon L.},
	year = {2017},
	month = {09},
	date = {2017-09-02},
	journal = {Journal of Statistics Education},
	pages = {137--144},
	volume = {25},
	number = {3},
	doi = {10.1080/10691898.2017.1395302},
	url = {\url{https://doi.org/10.1080/10691898.2017.1395302}},
	publisher = {Taylor \& Francis}
}

@article{ludlow2002,
	title = {Rethinking Practice: Using Faculty Evaluations to Teach Statistics},
	author = {Ludlow, Larry H.},
	year = {2002},
	month = {01},
	date = {2002-01-01},
	journal = {Journal of Statistics Education},
	volume = {10},
	number = {3},
	doi = {10.1080/10691898.2002.11910680},
	url = {\url{https://doi.org/10.1080/10691898.2002.11910680}},
	publisher = {Taylor \& Francis}
}

@article{boger2001,
	title = {The Benefit of Student-Generated Data in an Introductory Statistics Class},
	author = {Boger, Pam},
	year = {2001},
	month = {09},
	date = {2001-09-01},
	journal = {Journal of Education for Business},
	pages = {5--8},
	volume = {77},
	number = {1},
	doi = {10.1080/08832320109599663},
	url = {\url{https://doi.org/10.1080/08832320109599663}},
	publisher = {Routledge}
}

@article{gunderman2020,
	title = {Developing Lesson Plans for Teaching Spatial Data Management in Academic Libraries through a Lens of Popular Culture},
	author = {Gunderman, Hannah C.},
	year = {2020},
	month = {09},
	date = {2020-09-01},
	journal = {Journal of Map \& Geography Libraries},
	pages = {239--253},
	volume = {16},
	number = {3},
	doi = {10.1080/15420353.2021.1944948},
	url = {\url{https://doi.org/10.1080/15420353.2021.1944948}},
	publisher = {Routledge}
}

@article{brearley2018,
	title = {The {TSHS} Resources Portal: A Source of Real and Relevant Data for Teaching Statistics in the Health Sciences},
	author = {Brearley, Ann M. and Bigelow, Carol and Poisson, Laila M. and Grambow, Steven C. and Nowacki, Amy S.},
	year = {2018},
	date = {2018},
	journal = {Technology Innovations in Statistics Education},
	volume = {11},
	number = {1},
	doi = {10.5070/T5111034506},
	url = {\url{https://doi.org/10.5070/T5111034506}},
	langid = {en}
}

@article{bibby2003,
	title = {Rejoinder to {"}50 Years of Statistics Teaching in {E}nglish Schools: Some Milestones{"}, Holmes, P. Statistician},
	author = {Bibby, John},
	year = {2003},
	date = {2003},
	journal = {Journal of the Royal Statistical Society: Series D (The Statistician)},
	pages = {439--474},
	volume = {52},
	number = {4},
	langid = {en}
}

@inbook{hall2011,
	title = {Engaging Teachers and Students with Real Data: Benefits and Challenges},
	author = {Hall, Jennifer},
	year = {2011},
	date = {2011},
	publisher = {Springer Netherlands},
	pages = {335--346},
	doi = {10.1007/978-94-007-1131-0_32},
	address = {Dordrecht},
	langid = {en}
}

@inbook{cobb1992,
	title = {Teaching Statistics},
	author = {Cobb, George},
	year = {1992},
	date = {1992},
	pages = {3--43},
	series = {MAA notes},
	number = {22},
	publisher = {The Mathematical Association of American},
	langid = {en}
}

@article{braun2006,
	title = {Using thematic analysis in psychology},
	author = {Braun, Virginia and and Clarke, Victoria},
	year = {2006},
	month = {01},
	date = {2006-01-01},
	journal = {Qualitative Research in Psychology},
	pages = {77--101},
	volume = {3},
	number = {2},
	doi = {10.1191/1478088706qp063oa},
	url = {\url{https://doi.org/10.1191/1478088706qp063oa}},
	publisher = {Routledge}
}

@techreport{gaise2016,
  author      = {{GAISE College Report ASA Revision Committee}},
  title       = {Guidelines for Assessment and Instruction in Statistics Education College Report 2016},
  year        = {2016},
  url         = {http://www.amstat.org/education/gaise}
}

@inproceedings{young-saver2024,
  author       = {Young-Saver, Dash},
  title        = {We Went From 2\% to 42\% Passing {AP} Stats. Here's What We Learned},
  booktitle    = {2024 Electronic Conference on Teaching Statistics},
  year         = {2024},
  url          = {https://youtu.be/nII1hIhAY_E},
  note         = {Keynote}
}

@article{cetinkaya-rundel2022,
	title = {The 5Ws AND 1H OF TERM PROJECTS IN THE INTRODUCTORY DATA SCIENCE CLASSROOM},
	author = {Cetinkaya-Rundel, Mine and Dogucu, Mine and Rummerfield, Wendy},
	year = {2022},
	month = {07},
	date = {2022-07-04},
	journal = {Statistics Education Research Journal},
	pages = {4--4},
	volume = {21},
	number = {2},
	doi = {10.52041/serj.v21i2.37},
	url = {\url{https://doi.org/10.52041/serj.v21i2.37}},
	note = {Number: 2},
	langid = {en}
}

@article{evans2023,
	title = {Repurposing a Peer-Reviewed Publication to Engage Students in Statistics: An Illustration of Study Design, Data Collection, and Analysis},
	author = {Evans, Ciaran and Cipolli, William and Draper, Zakary A. and Binfet, John-Tyler},
	year = {2023},
	month = {09},
	date = {2023-09-02},
	journal = {Journal of Statistics and Data Science Education},
	pages = {236--247},
	volume = {31},
	number = {3},
	doi = {10.1080/26939169.2023.2238018},
	url = {\url{https://doi.org/10.1080/26939169.2023.2238018}},
	publisher = {Taylor \& Francis}
}

@article{weiland2024,
	title = {Culturally Relevant Data in Teaching Statistics and Data Science Courses},
	author = {Weiland, Travis and Williams, Immanuel},
	year = {2024},
	month = {09},
	date = {2024-09-01},
	journal = {Journal of Statistics and Data Science Education},
	pages = {256--271},
	volume = {32},
	number = {3},
	doi = {10.1080/26939169.2023.2249969},
	url = {\url{https://doi.org/10.1080/26939169.2023.2249969}},
	publisher = {Taylor \& Francis}
}

@article{whitehead2023,
	title = {Microbiome of Clothing Items Worn for a Single Day in a Non-Healthcare Setting},
	author = {Whitehead, Kelly and Eppinger, Jake and Srinivasan, Vanita and Ijaz, M. Khalid and Nims, Raymond W. and McKinney, Julie},
	year = {2023},
	month = {09},
	date = {2023-09},
	journal = {Microbiology Research},
	pages = {948--958},
	volume = {14},
	number = {3},
	doi = {10.3390/microbiolres14030065},
	url = {\url{https://doi.org/10.3390/microbiolres14030065}},
	publisher = {Multidisciplinary Digital Publishing Institute},
	langid = {en}
}

\end{document}